\documentclass[aps,prl,twocolumn,superscriptaddress,citeautoscript,showpacs]{revtex4}

\usepackage{graphicx}
\usepackage{amsmath}

\begin{document}

\title{Electronic and spin properties of Rashba billiards}

\author{J. Cserti}
\affiliation{Department of Physics of Complex Systems, E{\"o}tv{\"o}s
University, H-1117 Budapest, P\'azm\'any P{\'e}ter s{\'e}t\'any 1/A, Hungary}

\author{A. Csord\'as}
\affiliation{Department of Physics of Complex Systems, E{\"o}tv{\"o}s
University, H-1117 Budapest, P\'azm\'any P{\'e}ter s{\'e}t\'any 1/A, Hungary}
 
\author{U. Z\"ulicke}
\affiliation{Institute of Fundamental Sciences, Massey University, Private Bag
11~222, Palmerston North, New Zealand}  

\date{\today}

\begin{abstract}

Ballistic electrons confined to a billiard and subject to spin--orbit coupling
of the Rashba type are investigated, using both approximate semiclassical and
exact quantum--mechanical methods. We focus on the low--energy part of the
spectrum that has negative eigenvalues. When the spin precession length is
smaller than the radius of the billiard, the low--lying energy eigenvalues turn
out to be well described semiclassically. Corresponding eigenspinors are found
to have a finite spin polarization in the direction perpendicular to the
billiard plane.

\end{abstract}


\maketitle

Spin-dependent phenomena in semiconductor nanostructures have become the focus
of strong interest recently~\cite{spintronics-book,Wolf-review:cikk}. In
nonmagnetic systems, intriguing effects can arise from the presence of
spin--orbit coupling. Structural inversion asymmetry in semiconductor
heterostructures has been shown~\cite{roessler} to give rise to a spin
splitting of the same type as was discussed in an early paper by Rashba~\cite
{Rashba:cikk}. Its tunability by external gate voltages~\cite{nitta,thomas}
has motivated the theoretical design of a spin--controlled field--effect
transistor\cite{Datta-Das:cikk}. Novel spin properties arise from the interplay
between Rashba spin splitting and further confinement of two--dimensional
electrons in quantum
wires~\cite{hausler,Mireles-Kirczenow:cikk,Uli-1:cikk,thomas_wire}
or dots~\cite{dot1,Michele-q-dot:cikk,dot2,dot3,malshukov,zaitsev}. Spin-orbit
coupling has also been shown to affect the statistics of energy levels and
eigenfunctions as well as current distributions~\cite{Berggren:cikk,
Sadreev:cikk}.  

In this work, we study a {\em Rashba billiard}, i.e., non-interacting
ballistic electrons moving in finite 2D regions whose dynamics is affected by
the Rashba spin--orbit coupling. The Rashba spin--orbit coupling strength can
be conveniently measured in terms of a wave--number scale $2 k_{\text{so}}$,
which corresponds to the Fermi--wave--vector difference for the two spin--split
subbands. Typical values for the spin--orbit--induced spin precession length
$L_{\text{so}}=\pi/k_{\text{so}}$ are of the order of a few hundred
nanometers~\cite{spintronics-book}. The relevant parameter characterizing a
Rashba billiard of size $L$ is $k_{\text{so}} L$. Taking $L=10 \mu$m for a
typical size of quantum dots, the relevant parameter in Rashba billiards can
be as large as 70. Furthermore, the tunability of the Rashba spin--orbit
coupling strength is a convenient tool to induced changes of the billiard's
energy spectrum without applying external magnetic fields.

Below we present interesting features of the energy spectrum for Rashba
billiards, focusing especially on its negative energy eigenvalues. We will show
that the density of states (DOS) is singular at the bottom of the spectrum.
This singular behavior occurs independently of the billiard's shape and is
most striking if the Rashba parameter is large. We have found that for a
circular shape, the DOS has additional singularities at negative energies. We
obtain analytic results for their positions. Their corresponding eigenspinors
have a finite spin projection in the direction perpendicular to the billard
plane, which is the direct result of imposing hard--wall boundary conditions.

Our central quantity of interest is the Green's function for Rashba billiards.
Having obtained it, we can derive the density of states (DOS) $\varrho(E)$ 
and the smooth counting function $\bar{N}(E)$, i.e., the leading Weyl law~\cite
{Weyl:cikk,Baltes-Hilf_and_Brack:konyv}, plus correction terms to it. (A
discussion of these concepts for billiards without spin--orbit coupling can
be found in Refs.~\onlinecite{Kac:cikk,Balian-Bloch:cikkek,
Stewartson-Waechter:cikk,Berry-Howls:cikk,Uzy-Sieber:cikk}.)
We provide analytic results for circular Rashba billiards and give the first
two leading terms for arbitrary shapes based on the image method of Berry and
Mondragon \cite{Berry-neutrino:cikk}. The latter was developed for neutrino
billiards, which have two--component wave functions and are rather similar to
the Rashba billiards discussed here. Comparison of our analytical results
from the semiclassical treatment to the numerically calculated exact energy
levels of circular Rashba billiards demonstrate perfect agreement between the
two. An asymptotic expansion of spinor wave functions for negative--energy
eigenstates yields a finite spin polarization in the direction perpendicular
to the billiard, in contrast to the familiar result for a 2D plane.

In the one-band effective-mass approximation the Hamiltonian with Rashba
splitting in 2D is given by~\cite{bychkov}
\begin{subequations}
\begin{eqnarray}
\hat{H} &=& \frac{p_x^2 + p_y^2}{2m^*} + \frac{\alpha}{\hbar}\, \hat{U}
\quad ,\\  
\hat{U} &=& \sigma_x p_y - \sigma_y p_x \quad ,
\end{eqnarray}
\label{Rashba-Hamiltonian:eq}
\end{subequations}

\noindent
where $\sigma_x,\sigma_y$ are Pauli matrices. This Hamiltonian governs
the electron dynamics inside the billiard with 
Dirichlet boundary conditions at the perimeter. (See Ref. 
\cite{Bulgakov-Sadreev:cikk}.) $\pi/k_{\text{so}}$ is the spin--precession
length, which can be tuned independently of the system size 
\cite{Datta-Das:cikk,nitta,thomas}. In the absence of any lateral confinement,
the energy dispersion splits into two branches~\cite{bychkov}:
\begin{equation}
E (k_x,k_y) = \frac{\hbar^2}{2m^*}\, 
\left[{\left(k \pm  k_{\text{so}}\right)}^2 -  k_{\text{so}}^2\right], 
\label{dispersion:eq}
\end{equation}
where $k=\sqrt{k_x^2 + k_y^2}$. In the range $0< k< k_{\text{so}}$, one branch
has negative energies bounded from below by $-\Delta_{\text{so}}\equiv -\hbar^2
k_{\text{so}}^2/(2m^*)$. The spin splitting is a consequence of broken
spin-rotational invariance. The spin of energy eigenstates, which are labeled
by a 2D vector ${\bf k}$, is polarized perpendicularly to ${\bf k}$~\cite
{bychkov}. Hence, no common spin quantization axis for single--electron states
can be defined in the presence of spin--orbit coupling. At a given energy $E$,
two propagating modes exist whose wave vectors can be found from the dispersion
relation (\ref{dispersion:eq}): $k_{\pm} =k \mp k_{\text{so}}$, where $k=\sqrt
{\frac{2m^* E}{\hbar^2}+k_{\text{so}}^2}$. 

The Hamiltonian $\hat{H}$ commutes with the total angular momentum operator
$\hat{J}_z =-i\hbar \partial_\varphi + \frac{\hbar}{2}\,\sigma_z$, where
$\varphi$ is the polar angle. The eigenspinors $| \chi^{(\pm)}_m  \, \rangle$
corresponding to the two bands of the Hamiltonian (\ref{Rashba-Hamiltonian:eq})
are given, in the representation of polar coordinates $r,\varphi$, by
\begin{equation}
\langle  \, {\bf r} \, \mid \, \chi^{(\pm)}_m  \, \rangle = 
\left(\begin{array}{c} \pm\, J_m(k_{\pm} r) \\ 
J_{m+1}(k_{\pm} r)\, e^{i\varphi}
\end{array}
\right) \, e^{im \varphi} \quad .
\label{eigenspinor:eq}
\end{equation}
Here $J_m(x)$ is the Bessel function of integral order $m$ and  $E> -
\Delta_{\text{so}}$. Other independent solutions can be obtained when $J_m(x)$
are replaced by $Y_m(x)$, $H^{(1)}_m(x)$, or $H^{(2)}_m(x)$. For Rashba
billiards with arbitrary shape, the eigenstates can be expanded in the basis
(\ref{eigenspinor:eq}) using linear combinations of both $\pm$ spinor states.  

To proceed further, we need the free--space Green's function for the Rashba
Hamiltonian (\ref{Rashba-Hamiltonian:eq}). Using the fact that $\hat{U}^2 =
p_x^2 +p_y^2$ the operator $\hat{G}_{\infty} = {\left(E -\hat{H}\right)}^{-1}$
reads  
\begin{eqnarray}
\hat{G}_{\infty} &=& 
\frac{m^*}{\hbar^2} \, 
\frac{1}{k} \, \left[
\left(k_+ + \hat{U} \right){\left(k^2_+ 
-\frac{{\bf p}^2}{\hbar^2}\right)}^{-1}   
\right.  \nonumber \\ 
&& \left. + \left(k_- - \hat{U} \right)
{\left(k^2_- -\frac{{\bf p}^2}{\hbar^2}\right)}^{-1}
\right] .
\label{free-Green:eq}
\end{eqnarray}
Here $E$ can be a complex number. We note that, for negative energies, the
retarded Green's function contains incoming circular waves besides outgoing
waves.  

To satisfy the boundary conditions that $G({\bf r},{\bf r}^\prime)$ vanishes
at the boundary, the Green's function is decomposed into two parts:  $\hat{G}
= \hat{G}_\infty + \hat{G}_{\rm H}$, where the homogeneous part $\hat{G}_{\rm
H}$ fulfills $\left(E - \hat{H} \right )\, \hat{G}_{\text{H}} = 0$. This
latter Green's function is constructed, in the usual way, from the
eigenspinors (\ref{eigenspinor:eq}) as follows  
\begin{eqnarray}
\hat{G}_{\text{H}} &=& 
\sum_{m=-\infty}^\infty \left[ 
A_m \mid \chi^{(+)}_m  \, \rangle \, 
\langle \,  \chi^{(+)}_m  \, \mid  
+ B_m  \mid  \, \chi^{(-)}_m  \, \rangle  \, 
\langle  \, \chi^{(+)}_m  \, \mid  
\right.  \nonumber \\
&& \left. 
+ \,  C_m  \mid  \, \chi^{(+)}_m \,  \rangle \, 
\langle \,  \chi^{(-)}_m  \, \mid 
+ D_m  \mid \,  \chi^{(-)}_m  \, \rangle \,  
\langle \,  \chi^{(-)}_m \,  \mid  
\right].
\end{eqnarray}
The coefficients $A_m,B_m,C_m$ and $D_m$ should be chosen such that 
the total Green's function satisfies Dirichlet boundary conditions.
In general, one gets an infinite set of inhomogeneous linear equations
for the coefficients, which can be solved only numerically. 
The circular billiard is a special case for which  the coefficients can
be given analytically (not presented here) for any $m$.

From $\hat{G}$, the DOS can be obtained by $\varrho(E)=-\frac{1}{\pi}
\lim_{\eta \to 0^+}{\rm Im}{\rm Tr}\,\, \hat{G}(E+i\eta)$, where the trace
means the limit ${\bf r} \to {\bf r}^\prime$, integration of ${\bf r}$
over the area of the billiard, and the trace in spin space. The counting
function is defined by $N(E) = \int_{\infty}^E \, 
\varrho(E^\prime) d E^\prime$ and its smooth part $\bar{N}(E)$ requires
averaging over a small energy range around $E$.    

We now consider the circular Rashba billiard of radius $R$. Following the
ideas of the systematic method of Berry and Howls \cite{Berry-Howls:cikk}, we
have calculated the first few leading terms of $\bar{N}(E)$. (Details of the
lengthy calculation will be published elsewhere.) The result is: 
\begin{widetext}
\begin{equation}
\bar{N}(\varepsilon) = \left\{ \begin{array}{ll}
\frac{\varepsilon+2\varepsilon_{\text{so}}}{2}
-\sqrt{\varepsilon+\varepsilon_{\text{so}}} 
+ \frac{2}{\pi} \left[
\frac{\varepsilon}{\sqrt{\varepsilon + \varepsilon_{\text{so}}}}\, 
K\left(\sqrt{\frac{\varepsilon_{\text{so}}}
{\varepsilon + \varepsilon_{\text{so}}}} \right) 
- \sqrt{\varepsilon + \varepsilon_{\text{so}}} \, 
E\left(\sqrt{\frac{\varepsilon_{\text{so}}}
{\varepsilon + \varepsilon_{\text{so}}}} \right) 
\right], & 
\text{for}\,\,\, \varepsilon  > 0,  \\
\sqrt{\varepsilon_{\text{so}}}\sqrt{\varepsilon + \varepsilon_{\text{so}}} 
-\sqrt{\varepsilon + \varepsilon_{\text{so}}} 
-\frac{2\sqrt{\varepsilon_{\text{so}}}}{\pi} \, 
E\left(\sqrt{\frac{\varepsilon + \varepsilon_{\text{so}}}
{\varepsilon_{\text{so}}}} \right), &
 \text{for}\,\,\, -\varepsilon_{\text{so}} < \varepsilon  < 0,  
\end{array} \right.
\label{N_E-circular:eq}
\end{equation}
\end{widetext}
where the dimensionless energies 
$\varepsilon = 2m^*ER^2/\hbar^2$ and 
$\varepsilon_{\text{so}} = 2m^* \Delta_{\text{so}} R^2/\hbar^2 
= k_{\text{so}}^2 R^2$ have been introduced. Here $K(x)$ and $E(x)$ are
the complete elliptic integrals of the first and second kind, 
respectively, with the same definitions 
as in Ref.~\onlinecite{Gradshteyn-Ryzhik:konyv}.
The first term for both positive and negative energies is the
contribution from $\hat{G}_\infty$, while the remainder originates from
$\hat{G}_{{\rm H}}$.
We note that in a completely different context, namely for annular
ray-splitting billiards, a similar Weyl formula has been
calculated\cite{ray-splitting:cikk} involving also
elliptic integrals.    

For arbitrary shapes of Rashba billiards, we can also determine the area and
the perimeter terms of the smooth part of the counting function. The largest
contribution to $\bar{N}(E)$ comes from ${\rm Tr}\, \hat{G}_\infty$, which is
always proportional to the area $A$ of the billiard:
\begin{equation}
\!\!\!\!\bar{N}_1(E) =
\frac{A m^*}{2\pi \hbar^2} \! \left\{
\begin{array}{l} \!
\frac{E}{2} + \Delta_{\text{so}}, \,\, \text{for}\,\, E> 0, \\[1ex]
\!\!\! \sqrt{\Delta_{\text{so}}}\sqrt{E+\Delta_{\text{so}}}, 
\, \text{for}\,-\Delta_{\text{so}}\!\!< \! E \! < 0. 
\end{array} \right.  
\label{leading_N:eq}
\end{equation}
It follows directly from Eq.~(\ref{leading_N:eq}) that, for negative energies,
the DOS shows a $1/\sqrt{E+\Delta_{\text{so}}}$ singularity at the bottom
of the spectrum $E \to -\Delta_{\text{so}}$. The perimeter term can be derived
from the generalization of the image method of
Ref.~\onlinecite{Balian-Bloch:cikkek} using only the free space Green's
function. The calculation is very much similar to that applied by Berry and
Mondragon~\cite{Berry-neutrino:cikk} for neutrino billiards and yields 
\begin{equation}
\bar{N}_2(E) = 
-\frac{\cal{L}}{2\pi}\, \sqrt{\frac{2m^*}{\hbar^2}}\, 
\sqrt{E+ \Delta_{\text{so}}},
\label{perim_N:eq}
\end{equation}
valid for all energies $E>-\Delta_{\text{so}}$. Here $\cal{L}$ is the
perimeter length of the billiard. The minus sign is a consequence of Dirichlet
boundary conditions. For zero spin--orbit coupling, $\bar{N}_1(E)+\bar{N}_2(E)$
coincides with the previously derived result for 2D
billiards~\cite{Baltes-Hilf_and_Brack:konyv} (apart from a factor 2 due to
spin). The area term (\ref{leading_N:eq}) can alternatively be derived from
the classical phase-space integral in the underlying classical approach.
However, the classical dynamics of electrons in Rashba billiards is described
by {\em two} Hamiltonians \cite{Pletyukhov-Brack:cikk}, which  are reminiscent
of the two dispersion branches (\ref{dispersion:eq}). The constant-energy
surfaces in phase space are different for the two Hamiltonians, yielding
different contributions to the classical phase-space integral. A simple
calculation gives then Eq.~(\ref{leading_N:eq}). 
  
\begin{figure}[t]
\includegraphics[scale=0.5]{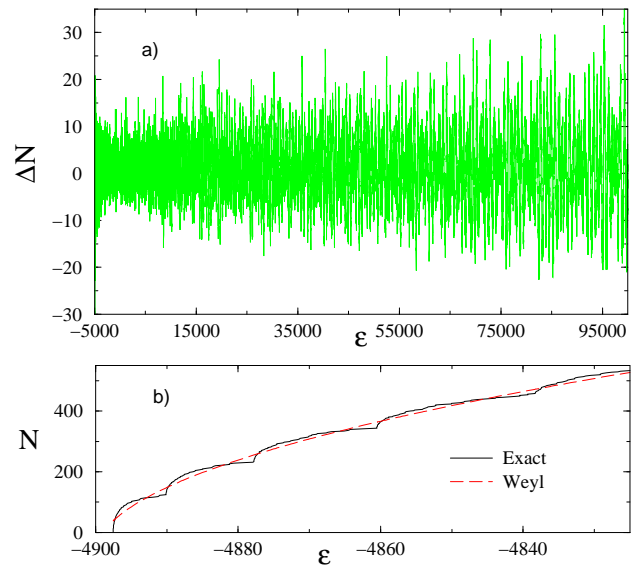}
\caption{In panel a) the difference $\Delta N$ between 
the exact counting function and $\bar{N}(\varepsilon) $
from Eq.~(\ref{N_E-circular:eq}) is plotted
for $\sqrt{\varepsilon_{\text{so}}}=k_{\text{so}} R =70$.
In panel b) the exact counting function (solid line) and
$\bar{N}(\varepsilon) $ (dashed line) are shown.
In both panels dimensionless energies $\varepsilon = 2m^*ER^2/\hbar^2$
are used.  
\label{N_E:figure}}
\end{figure}

To get better agreement between the numerically obtained exact counting
function and $\bar{N}(E)$ one should calculate further terms besides
$\bar{N}_1(E)$ and $\bar{N}_2(E)$ for large $k_{\text{so}}$. This motivated us
to consider more corrections in Eq.~(\ref{N_E-circular:eq}) involving elliptic
integrals. For the case of circular Rashba billiards, the Schr{\"o}dinger
equation is separable in polar coordinates. The resulting radial equation for
both spinor components leads to the secular
equation~\cite{magneses-dot_SO-exact:cikk} $J_m(k_+R)J_{m+1}(k_-R) +
J_m(k_-R) J_{m+1}(k_+R) = 0$, where $m$ is an integer. This equation is
invariant under the change $m \rightarrow -m-1$ (Kramers degeneracy). Formal
solutions of the secular equation having zero wave vector are excluded since
the corresponding wave functions vanish everywhere inside the billiard. We
obtain the exact $N(\varepsilon)$ from our solutions of the secular equation
for different $m$. In Fig.~\ref{N_E:figure}a, we plot the difference $\Delta
N=N(\varepsilon)-\bar{N}(\varepsilon)$ as a function of the dimensionless
energy $\varepsilon$.
The difference fluctuates around zero, which shows we did not miss levels
(the mean value of $\Delta N$ is a sensitive test for missing levels, 
see e.g.,  Ref.~\onlinecite{Schmit:cikk_Csordas:cikk}). 
Fig.~\ref{N_E:figure}a shows data for approximately 54570 levels.
Without correction terms in Eq.~(\ref{N_E-circular:eq}) with elliptic
integrals, $\Delta N$ would increase monotonically on average, and would
predict $\approx 70$ missing levels in the energy range plotted.

\begin{figure}[hbt]
\includegraphics[scale=0.45]{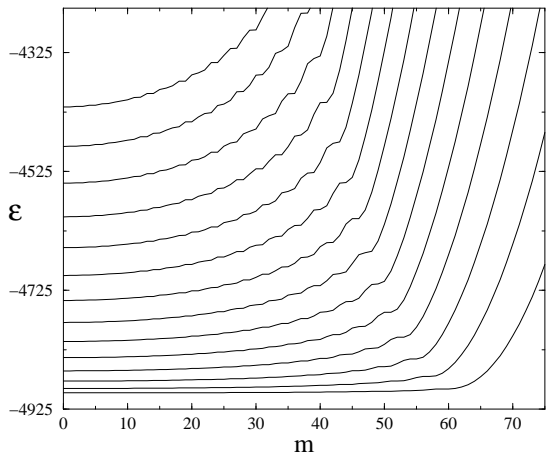}
\caption{The energy levels (in units of $2m^*R^2/\hbar^2$) of circular
Rashba billiards 
as functions of $m$. For a given $m$ levels are ordered and
first, second, etc.\ energies are connected as $m$ varies.
\label{E_m:figure}}
\end{figure}

In Fig.~\ref{N_E:figure}b, the exact counting function is shown together  
with the Weyl formula (\ref{N_E-circular:eq}) for negative energies near
the bottom of the spectrum $ -\varepsilon_{\text{so}}$. The overall 
agreement is good but the exact $N(\varepsilon)$ shows an additional 
rounded step structure at certain energies $\varepsilon_n$. 
This feature shows up only for negative energies, although for 
larger energies this is less pronounced.  
The step structure results in large deviations $\Delta N$ at energies
$\varepsilon_n$ and concomitant large peaks in the DOS.
To see the reason for this behavior,
it is useful to plot the energy levels as functions of $m$, as shown 
in Fig.~\ref{E_m:figure}.   
The curves in the figure start almost horizontally at $\varepsilon_n$, 
$n=1,2,\ldots$ resulting in large peaks in the DOS at the same energies.

Using Hankel's asymptotic expression for Bessel functions with large
argument~\cite{Abramowitz-Stegun:konyv}, we were able to derive the
energy dispersion accurately in next--to--leading order:
\begin{subequations}\label{e_sing:eq}
\begin{eqnarray} 
\varepsilon_{m,n}&=&{(\frac{n\pi}{2})}^2-\varepsilon_{\text{so}}+\delta_{m,n}
\quad,\\ 
\delta_{m,n}&=&\left(\frac{n\pi}{2}\right)^2 \frac{2 m+1}{\varepsilon_{\text
{so}}} \left[m+1 + (-1)^n \cos(2\sqrt{\varepsilon_{\text{so}}})\right] \quad ,
\end{eqnarray}
\end{subequations}
valid only for negative energies and $n=1,2,\dots,{\mathrm{int}}(2
\sqrt{\varepsilon_{\text{so}}}/\pi)$. For small $m,n$ the above expression
agrees excellently with the numerics (e.g., $\varepsilon_{0,1}$ is accurate up
to 8 digits for $\varepsilon_{\text{so}}=70$). It is straightforward to obtain
corresponding spinor eigenstates and calculate their expectation value for the
$z$ component of spin. Similar to the case of Rashba--split eigenstates in
rings~\cite{Uli_persistent:cikk}, but in contrast to that of quantum
wires~\cite{hausler,Uli-1:cikk}, it turns out to be finite. We find
\begin{equation}
\big\langle\sigma_z\big\rangle_{m,n}=-\frac{\left(\frac{n\pi}{2\sqrt
{\varepsilon_{\text{so}}}}\right)^2 \cos\left(2\sqrt{\varepsilon_{\text
{so}}}\right)}{(2 m + 1)\cos\left(2\sqrt{\varepsilon_{\text{so}}}\right) +
2 (-1)^{n+m} \frac{\varepsilon_{\text{so}}-\left(\frac{n\pi}{2}\right)^2}
{\sqrt{\varepsilon_{\text{so}}}}} .
\end{equation}

The Schr{\"o}dinger equation (including boundary conditions) 
for circular Rashba billiards is separable
in polar coordinates, thus integrable. Hence, all the level statistics
should be Poissonian (see e.g.~Ref.~\onlinecite{Q-billiard:konyv}).
Indeed, we have found that the nearest--neighbor 
level--spacing distribution $P(s)$ is Poissonian (not shown here).  
For other shapes, the spin-orbit coupling destroys the integrability.
Random Matrix Theory predicts the statistics to be that of the Gaussian
Orthogonal Ensemble (GOE) due to time reversal symmetry
\cite{Mireles-Kirczenow:cikk}. Some other intermediate distribution 
\cite{Berggren:cikk} has been found for rectangular shape and for small
$k_{\text{so}}$, reflecting the fact that the rectangular billiard
without SO coupling is integrable.     

Finally, a few interesting open theoretical problems are listed. 
The Weyl formula is essential to develop a periodic orbit theory 
for Rashba billiards. (For normal billiards, see Brack and Bhaduri's book 
in Ref.~\onlinecite{Baltes-Hilf_and_Brack:konyv}, and a theory in
case of harmonically confined Rashba systems is given in
Ref.~\onlinecite{Pletyukhov-Brack:cikk}.) The Green's function method
presented in this work would be a suitable starting point to calculate
observables such as the magnetization~\cite{Silvia-Rocca-1:cikk} or 
persistent currents~\cite{Uli_persistent:cikk} in Rashba billiards. 

In conclusion, we have presented a study of electron billiards with
spin--dependent dynamics due to Rashba spin splitting.
Semiclassical results for the spectrum agree well with exact
quantum calculations. We find interesting properties of negative--energy
states, including a finite spin projection in the out--of--plane
direction.

This work was supported by the Hungarian Science Foundation OTKA  T034832 and
T038202. U.~Z. gratefully acknowledges funding from the MacDiarmid Institute
for Advanced Materials and Nanotechnology.

\end{document}